# Experimental nuclear charge density and theoretical description of the above-barrier light heavy-ion fusion process

I. I. Gontchar(a), M. V. Chushnyakova(b)

*(a)Physics and Chemistry Department, Omsk State Transport University, Omsk, Russia*
*(b)Physics Department, Omsk State Technical University, Omsk, Russia*

**Abstract:** Theoretical modeling of nucleus-nucleus collision often is based on the nucleus-nucleus potential. One of the advanced methods for constructing this potential is the semi-microscopical double-folding model with the M3Y-Paris NN-forces. Proton and neutron densities are significant ingredient of this model. Correct nucleon density (ND) must reproduce experimental nuclear charge density (NCD). However, those who deal with modeling the fusion process, typically, disregard this circumstance. We aim to achieve good description of both the charge nucleon density and the above-barrier fusion cross sections of even-even light nuclei with $Z = N$. We consider several versions of NDs available in the literature and construct our own approximation for the ND of even-even spherical nuclei 12C, 16O, 40Ca which is abbreviated as FE-density (Fermi+exponential). We carefully compare the NCDs resulting from different versions of NDs with the experimental NCSs. After finding the nucleus-nucleus potential using the double-folding model with the density dependent M3Y-Paris NN-forces and FE densities we evaluate the above-barrier fusion cross sections for five reactions, 12C+12C, 12C+16O, 16O+16O, 16O+40Ca, and 40Ca+40Ca, where the experimental data are available. The cross sections are calculated using two approaches: a) the barrier penetration model and b) the Trajectory Model with surface friction (TM). To find the transmission coefficients for TM, the Langevin equations are employed. For all considered reactions, our TM typically reproduces the above-barrier experimental cross sections within 10-15%. The only adjustable parameter of the model, the optimal friction strength K_Rm, was found to be about 90 zs/GeV for light reactions 12C+12C, 12C+16O, 16O+16O and about 15 zs/GeV for heavier reactions 16O+40Ca and 40Ca+40Ca. The latter findings are in reasonable agreement with the systematics found earlier. Thus, the FE-recipe allows reproducing simultaneously with good accuracy both the charge nucleon density and the above-barrier fusion cross sections for five reactions involving 12C, 16O, 40Ca nuclei.

**Keywords:** nuclear charge densities, nucleon densities, heavy-ion fusion, surface friction, double-folding model

## I. INTRODUCTION

Theoretical modeling of the nucleus-nucleus collision is useful since it reduces the experimental labor and expenses. Moreover, this modeling provides us with understanding the mechanism of nucleus-nucleus interaction. For modeling the fusion process of two nuclei, the time-dependent Hartree–Fock approach [1-4] is often considered as one of the most advanced methods. However, there are two shortcomings of this approach: (i) it is very computer-time consuming and (ii) it is difficult to account for fluctuations of the nucleus shape. Therefore, in the literature simplified dynamical approaches are widely used like the quantum diffusion approach [5-8], the coupled-channels method [9-12] or the trajectory model with the surface friction [13-16].

In all these simplified approaches, a crucial role is played by the nucleus-nucleus effective interaction potential, consisting of the Coulomb $U_C$, nuclear $U_n$, and centrifugal $U_{rot}$ terms. The nuclear term of this potential is the least defined quantity. In many works, this term is calculated using the semi-microscopical double folding model with the frozen densities [5, 8, 17-22]. A simplest formula for this model reads

$$U_n(R) = \int d\vec{r}_1 \int d\vec{r}_2 \rho_{A1}(r_1) v_{NN}(|\vec{R} - \vec{r}_2 + \vec{r}_1|) \rho_{A2}(r_2). \tag{1}$$

Here $\rho_{A1}$ ($\rho_{A2}$) is the point nucleon projectile (target) density, $v_{NN}$ is the effective nucleon-nucleon interaction, $\vec{R}$ corresponds to the projectile-target center-to-center distance, $\vec{r}_1$ and $\vec{r}_2$ are the radius-vectors of the interacting points of the projectile and target nuclei, respectively. Note that in the present work, we consider collisions of two spherical nuclei; the nucleon density is the sum of the proton and neutron densities, $\rho_{Ai} = \rho_{Zi} + \rho_{Ni}$ ($i = 1,2$). For evaluating the Coulomb nucleus-nucleus interaction energy $U_C(R)$, one uses Eq. (1) changing appropriately the densities and nucleon-nucleon interaction.

The Nucleon Densities $\rho_{A1(2)}$ (NDs) come to Eq. (1) from different sources which are external with respect to the double folding approach. The problem is, however, that the NDs are not the observable quantities. Although recently some experimental information appeared on the neutron density [23, 24], the usual way to explore experimentally the interior of the nucleus is to irradiate it with electrons; thus, the measured quantity is the Nuclear Charge Density (NCD) [25-27]. For the collision of spherical nuclei, the NCD is not considered to be of crucial importance when calculating the Coulomb term of the nucleus-nucleus potential [18]. However, for consistency, the theoretical NDs used in Eq. (1) must provide the nuclear charge densities in agreement with the experimental NCDs as, we hope, it is in nature.

The idea of the present work is to make use of the experimental NCDs for obtaining the correct NDs which are then implemented in the double folding potential. Three spherical even-even nuclei with $N = Z$ are employed: $^{12}$C, $^{16}$O, and $^{40}$Ca. Although there is some experimental information on the deformed shape of $^{12}$C (see, e.g. [28, 29] and references therein), in many works this nucleus is considered as a spherical one [14, 20, 21, 30, 31]. Moreover, it is well established that, at the above barrier collision energies, the structure effects like nuclear deformations do not significantly manifest themselves [32]. Therefore, we believe it is reasonable approximation to consider $^{12}$C as a spherical nucleus in our present calculations.

In collision of light heavy ions, the capture of the reagents into the orbital motion results in their fusion into a compound nucleus. That is why we use the term "fusion" although in fact we calculate just capture cross sections. Following [21] we assume that $\rho_Z = \rho_N$ for these nuclei and propose a novel approach for parametrizing the NDs. These densities are used for evaluating the above-barrier capture cross-sections for $^{12}$C+$^{12}$C, $^{16}$O; $^{16}$O+$^{16}$O, $^{40}$Ca; and $^{40}$Ca+$^{40}$Ca reactions.

The paper is organized as follows. In Section II, we discuss the relation between the nuclear charge density and nucleon density as well as the different versions of ND available in the literature. In Section III, a novel parametrization for the nucleon density is described and the calculated NCDs are compared with the experimental ones. The models used for calculating the fusion cross sections are considered in Section IV. Theoretical fusion cross-sections are compared with the experimental ones in Section V. In Section VI, conclusions are formulated.

## II. NUCLEAR CHARGE DENSITY AND NUCLEON DENSITY

The experimental NCDs are approximated using the Sum of Gaussians (SOG) analysis [26]:

$$\rho_{q\,exp}(r) = \sum_i A_i \left\{ \exp\left[-\left(\frac{r-R_i}{\gamma}\right)^2\right] + \exp\left[-\left(\frac{r+R_i}{\gamma}\right)^2\right] \right\} \quad (2)$$

with

$$A_i = \frac{ZQ_i}{2\pi^{3/2}\gamma^3(1+2R_i^2/\gamma^2)} \quad (3)$$

where

$$\sum_i Q_i = 1. \quad (4)$$

The coefficients $Q_i$, as well as the positions $R_i$ and width $\gamma$ of the Gaussians are tabulated in [26].

Provided one knows the proton density $\rho_Z$, the NCD in a spherical nucleus, $\rho_q(r)$, as a function of the distance from its center, $r$, is calculated using the convolution method [33]

$$\rho_q(r) = \int d\vec{r}_p \rho_Z(r_p) f_p(|\vec{r}-\vec{r}_p|). \quad (5)$$

Here $\vec{r}_p$ denotes the radius-vector of the proton center of mass; $f_p$ is the charge distribution inside proton.

For this distribution as function of $l = |\vec{r}-\vec{r}_p|$, we use the exponential distribution [34]

$$f_{pe}(l) = \frac{1}{\pi s_{pe}^3} \exp\left(-\frac{2\sqrt{3}l}{R_{qp}}\right) \quad (6)$$

where $R_{qp}$ is the experimental value of the root mean square proton charge radius $R_{qp} = 0.8783$ fm [27]. Although in recent experiments [35, 36] this value has been found by 4% smaller, in the present work we use the one from [27].

In the literature one finds several options for the NDs. In [37], theoretical densities calculated within the Hartree-Fock-Bogolubov approach are approximated by the 2pF profiles with different radius and diffuseness parameters for protons (Z) and neutrons (N)

$$\rho_{i2pF}(r) = \frac{\rho_{i2pFC}}{1+\exp\left[\frac{r-R_{0i2pF}}{a_{i2pF}}\right]}. \quad (7)$$

Here $i = Z, N$, $R_{0i2pF}$ and $a_{i2pF}$ are half-central density radius and diffuseness parameters, respectively, $\rho_{i2pFC}$ comes from the normalization condition. However, for $^{12}$C this type of density profile is absent in [37].

For $^{12}$C and $^{16}$O, one finds in [21] the Gaussian-like densities

$$\rho_{AG}(r) = 2\rho_{NG}(r) = 2\rho_{ZG}(r) = \frac{4}{\pi^{\frac{3}{2}} s_{AG}^3}\left(1+\frac{F_{AG}r^2}{s_{AG}^2}\right)\exp\left(-\frac{r^2}{s_{AG}^2}\right) \quad (8)$$

with $s_{AG} = 1.5840$ fm, $F_{AG} = 4/3$ for $^{12}$C and $s_{AG} = 1.7410$ fm, $F_{AG} = 2$ for $^{16}$O. However, this type of density profile is not known to us for $^{40}$Ca.

In [8], an effective analytical method for evaluating the double-folding integrals was developed. For this aim for the nucleon densities a symmetrized Woods-Saxon function was used and applied systematically in work by Sargsyan et al. [5]. This function in our case reads

$$\rho_{AS}(r) = 2\rho_{NS}(r) = 2\rho_{ZS}(r) = \frac{\rho_{SC}\sinh(R_{0S}/a_S)}{\cosh(R_{0S}/a_S) + \cosh(r/a_S)} \quad (9)$$

Here $R_{0S} = r_{0S}A^{1/3}$, $r_{0S}$=1.15 fm, $a_A$=0.53 fm for $^{12}$C, $^{16}$O and 0.55 fm for $^{40}$Ca. We find $\rho_{SC}$ from the normalization condition.

### III. A NOVEL PARAMETRIZATION FOR THE NUCLEON DENSITIES

We propose a novel algorithm for finding the NDs based on the experimental NCDs. We call it Fermi+exponential and denote it as FE:

$$\rho_{AFE}(r) = 2\rho_{ZFE}(r) = 2\rho_{NFE}(r) = \begin{cases} \rho_{FEC}[1 + \exp\{(r - R_{0FE})/a_{AF}\}]^{-1} & at\ r < R_{0FE}, \quad (10) \\ 0.5\rho_{FEC}\exp\{(R_{0FE} - r)/a_{AE}\} & at\ r \geq R_{0FE}. \quad (11) \end{cases}$$

First, we approximate the ND by the Fermi profile (Eq. (10)). The values of $R_{0FE}$ and $a_{AF}$ are varied to reach a good agreement with the inner part of the experimental NCD. Then, for $r \geq R_{0FE}$, the ND is approximated by an exponential function (Eq. (11)). At $r = R_{0FE}$, Eqs. (10) and (11) result in $0.5\rho_{FEC}$. The diffuseness of the exponential tail, $a_{AE}$, is varied to fit the tail of the NCD, then the constant $\rho_{FEC}$ is found from the normalization condition:

$$A = \int d\vec{r}\rho_{AFE}(r). \quad (12)$$

The parameters $R_{0FE}$, $a_{AF}$, $a_{AE}$ providing the best agreement with the experimental NCDs are presented in Table I.

TABLE I. The experimental rms charge radii $r_{q\ exp}$ with their errors [27], the fractional differences for the radii $\xi_{rG}$, $\xi_{r2pF}$, $\xi_{rS}$, $\xi_{FE}$ and the average differences for the densities $\xi_{\rho G}$, $\xi_{\rho 2pF}$, $\xi_{\rho S}$, $\xi_{\rho E}$, the parameters resulting from FE algorithm $R_{0FE}$, $a_{AF}$, $a_{AE}$, the energy of the second excited state $E_2$.

| Nucleus | $r_{q\ exp}$ (fm) | exp err (fm) | $\xi_{rG}$ (%) | $\xi_{r2pF}$ (%) | $\xi_{rS}$ (%) | $\xi_{rFE}$ (%) | $\xi_{\rho G}$ (%) | $\xi_{\rho 2pF}$ (%) | $\xi_{\rho S}$ (%) | $\xi_{\rho FE}$ (%) | $R_{0FE}$ (fm) | $a_{AF}$ (fm) | $a_{AE}$ (fm) | $E_2$ (MeV) |
|---|---|---|---|---|---|---|---|---|---|---|---|---|---|---|
| $^{12}$C | 2.4702 | 0.0022 | 0.9 | n/a | 19.9 | -0.3 | 21.3 | n/a | 371 | 11.0 | 2.50 | 0.27 | 0.42 | 7.7 |
| $^{16}$O | 2.6991 | 0.0052 | 2.0 | n/a | 14.9 | 1.6 | 18.4 | n/a | 129 | 7.9 | 2.80 | 0.25 | 0.47 | 6.1 |
| $^{40}$Ca | 3.4778 | 0.0012 | n/a | 0.8 | 8.4 | -0.5 | n/a | 18.3 | 55.2 | 12.7 | 3.75 | 0.62 | 0.54 | 3.7 |

The NCDs resulting from $\rho_{AS}(r)$, $\rho_{AG}(r)$, $\rho_{Z2pF}(r)$ (i.e. $\rho_{qS}(r)$, $\rho_{qG}(r)$, $\rho_{q2pF}(r)$) as well as from the novel algorithm (FE algorithm henceforth, $\rho_{qFE}(r)$) are compared with the experimental NCDs in Figs. 1, 2 and 3 for $^{12}$C, $^{16}$O, and $^{40}$Ca, respectively. For $^{12}$C and $^{16}$O nuclei the Gaussian ND-profile results in good agreement with $\rho_{qexp}(r)$ in the nuclear interior, whereas for the tail $\rho_{qG}(r)$ underestimate the data. The FE-algorithm provide a somewhat poorer agreement with $\rho_{qexp}(r)$ in the interior but much better reproduction of the data in the tail. The NDs $\rho_{qS}(r)$ disagree with $\rho_{qexp}(r)$ for all values of $r$.

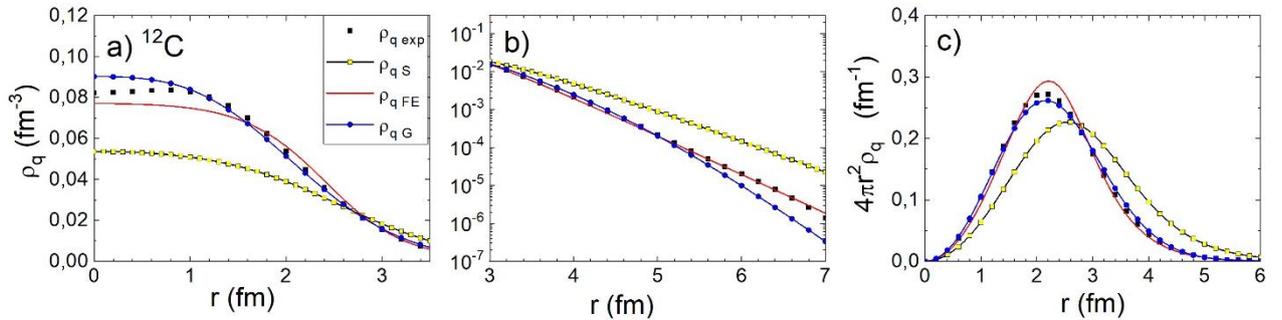

**Fig. 1.** (color online) The nucleus charge densities in linear (a) and logarithmic (b) scales and the "radial" charge density $4\pi r^2 \rho_q(r)$ (c) for $^{12}$C. Black squares (see Eq. (2)) indicate the experimental density, light squares stand for Sargsyan density, lines without symbols denote FE densities, and the line with solid circles represent Gaussian density.

For $^{40}$Ca (see Fig. 3, the Gaussian ND profile absents, and we compare with $\rho_{qexp}(r)$ the charge density resulting from 2pF ND instead. The charge densities $\rho_{q2pF}(r)$ and $\rho_{qFE}(r)$ are in good agreement with each other. For the interior they reproduce $\rho_{qexp}(r)$ well, whereas for the very tail the agreement worsens. The $\rho_{qS}(r)$ again significantly deviates from the data.

For the quantitative measure of the agreement between the theoretical and experimental NCDs, we use two quantities. The first one, the fractional difference between the root mean square radii, $\xi_{ru}$, reads

$$\xi_{ru} = {r_{q\,u}}/{r_{q\,exp}} - 1. \tag{13}$$

The second quantity is the average fractional difference between the NCDs, $\xi_{\rho u}$:

$$\xi_{\rho u} = \frac{1}{Q}\sum_{i=1}^{Q}\left|1 - \frac{\rho_{q\,u}(r_i)}{\rho_{q\,exp}(r_i)}\right|. \tag{14}$$

Here the summation is limited by the range of the experimental NCDs [26]. In Eqs. (13), (14) $u = G$, FE, 2pF, or S.

The results of the quantitative comparison are presented in Table I. One sees that for the nuclear charge density the FE algorithm provides significantly better agreement with the experimental values than the others NDs do. Yet the deviation $\rho_{qS}$ from the experimental NCDs is the most striking.

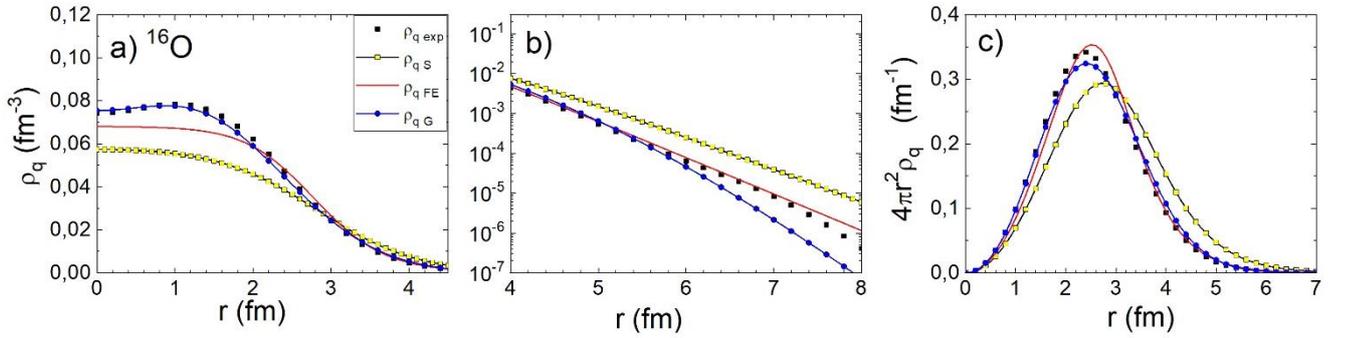

**Fig. 2.** (color online) Same as in Fig. 1 but for $^{16}$O.

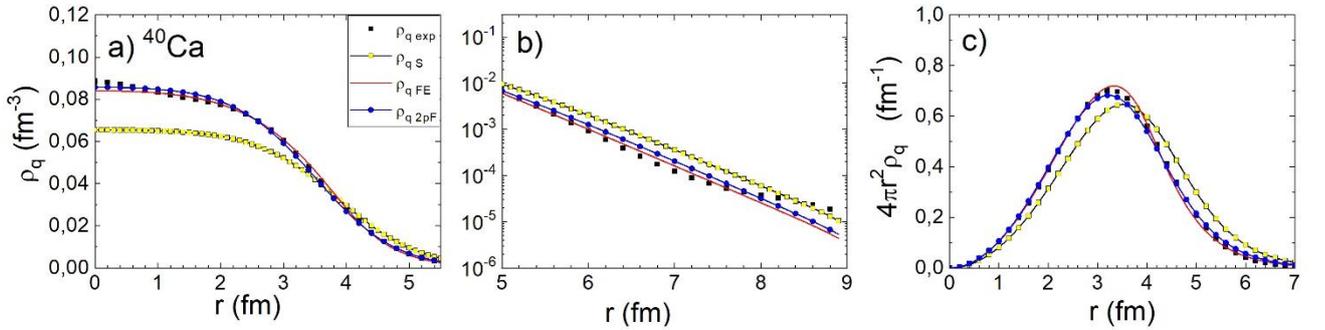

**Fig. 3.** (color online) Same as in Figs. 1, 2 but for $^{40}$Ca. The line with solid circles represents 2pF density.

### IV. MODELS FOR FUSION CROSS SECTIONS

Of four considered versions of the densities only $\rho_{AS}(r)$ and $\rho_{AFE}(r)$ are available for all nuclei $^{12}$C, $^{16}$O, and $^{40}$Ca. However, the nucleon densities $\rho_{AS}(r)$ of Refs. [5, 8, 19] fail to produce the charge densities in agreement with the experiment. Moreover, in that works, the Migdal Skyrme-type NN forces were used. That is why in the rest of the paper all calculations are performed only with FE-densities. Here we provide only a brief description of the options employed; all the details about the version of the double-folding model utilized in the present work can be found in [13, 15].

The potential energies for the reactions under consideration are calculated within the framework of the double folding model (see Eq. (1)) with the frozen densities $\rho_{AFE}(r)$. For the effective nucleon-nucleon forces, $v_{NN}$, the M3Y-Paris parametrization is applied. The nuclear term of the nucleus-nucleus potential consists of the direct and exchange parts. For the latter, the option with the finite range *NN*-forces is employed. Both direct and exchange parts are density dependent. There are about 10 parametrizations of this dependence [21, 31]; we use the DD2-option according to Table II of Ref. [38]. It is known in the literature that the version with the finite-range exchange term and with this density dependence of $v_{NN}$ successfully reproduces the nuclear matter equilibrium density, binding energy, and incompressibility as well as the high precision above-barrier fusion heavy-ion cross-sections.

Information on the reactions under consideration is comprised in Table II. One sees that for reactions $^{12}$C +$^{12}$C, $^{16}$O and $^{16}$O +$^{16}$O, $^{40}$Ca, the calculated s-wave barrier energies $B_0$ are in good agreement with the experimental ones. For reaction $^{40}$Ca+$^{40}$Ca, the agreement is somewhat worse. However, one should remember that what is called "experimental barrier energy" is based not solely on the measured fusion cross sections but on a model used for the analysis of these cross sections.

In the present paper, the capture cross-sections are calculated by means of the standard quantum mechanical formulas (see e.g. [39], Eqs. (1.38) and (1.143)):

$$\sigma_{th} = \frac{\pi \hbar^2}{2 m_R E_{c.m.}} \sum_{L=0,1,2,\ldots}^{L_{max}} (2L+1) T_L \qquad (15)$$

for $^{12}$C+$^{16}$O and $^{16}$O+$^{40}$Ca reactions and

$$\sigma_{th} = \frac{\pi \hbar^2}{m_R E_{c.m.}} \sum_{L=0,2,4\ldots}^{L_{max}} (2L+1) T_L. \qquad (16)$$

for the other reactions. Here $E_{c.m.}$ is the collision energy in the center-of-mass frame; $m_R = m_n A_1 A_2/(A_1 + A_2)$; $L$ denotes the angular momentum in units of $\hbar$; $L_{max}$ is the maximal angular momentum above which the transmission coefficient becomes equal to zero; $m_n$ stands for the mass of nucleon.

TABLE II. For five reactions under consideration, the following quantities are presented: the s-wave barrier radius $R_{b0}$ and width $\hbar\Omega_{b0}$; $B_Z$ (see Eq. (23)); the s-wave barrier heights $B_0$ calculated in the present work; the s-wave experimental barrier height $B_{0exp}$ (with its error) and the corresponding references; the references to the experimental cross sections; the value $E_{cut}$ discussed in Sec. V; the number of data points used for the detailed quantitative analysis $v$; the friction strength $K_{Rm}$ (with its error) providing the least value of $\chi^2$, $\chi_m^2$ (see Eq. (21)), $E_{cut} = B_0 + E_2$.

| Reaction | $R_{b0}$ (fm) | $\hbar\Omega_{b0}$ (MeV) | $B_Z$ (MeV) | $B_0$ (MeV) | $B_{0exp}$(err) (MeV) | Source of $B_{0exp}$ | Source of $\sigma_{exp}$ | $E_{cut}$ (MeV) | $v$ | $K_{Rm}$(err) (zs·GeV$^{-1}$) | $\chi_m^2$ |
|---|---|---|---|---|---|---|---|---|---|---|---|
| R1 $^{12}$C+$^{12}$C | 7.73 | 2.17 | 7.87 | 6.22 | 5.8(0.3) | [40] | [41] | 13.9 | 27 | 77(7) | 2.8 |
|  |  |  |  |  | 6.17(0.10) | [42] | [40] |  | 45 | 88(10) | 2.9 |
| R2 $^{12}$C+$^{16}$O | 8.12 | 2.15 | 9.99 | 7.90 | 7.7(0.4) | [41] | [41] | 14.0 | 25 | 85(10) | 1.6 |
|  |  |  |  |  | 7.69(0.10) | [43] |  |  |  |  |  |
| R3 $^{16}$O+$^{16}$O | 7.99 | 2.33 | 12.71 | 10.72 | 11.2(0.6) | [41] | [44] | 16.8 | 48 | 44(8) | 7.1 |
| R4 $^{16}$O +$^{40}$Ca | 9.29 | 2.46 | 26.97 | 23.04 | 23.7(1.0) | [45] | [46] | 26.7 | 63 | 18(7) | 0.4 |
| R5 $^{40}$Ca +$^{40}$Ca | 10.04 | 2.55 | 58.55 | 53.30 | 50.6(2.8) | [41] | [47] | 57.0 | 4 | 13(11) | 0.9 |
|  |  |  |  |  | 51.5(0.5) | [48] | [48] |  | 11 | 7(6) | 2.5 |

In the present work, the transmission coefficients are evaluated using two options. As the first one, the Barrier Penetration Model (BPM) within the framework of the parabolic approximation is applied:

$$T_{L\,BPM} = \{1 + \exp[2\pi(U_{bL} - E_{c.m.})/(\hbar\Omega_{bL})]\}^{-1}. \qquad (17)$$

Here $U_{bL}$ and $\Omega_{bL} = \sqrt{C_{2bL}/m_q}$ denote the barrier height and curvature calculated for the $L$-th partial wave (see definition of $m_q$ below). The cross-sections evaluated by means of this option are denoted as $\sigma_{BPM}$. They approximately correspond to the upper limit of the theoretical cross-sections $\sigma_{th}$ obtained dynamically because accounting for dissipation inhibits the fusion process.

The second option for getting the transmission coefficient is the Trajectory Model with surface friction (TM) [13, 49]. The physical picture of the TM is similar to that of Ref. [50]. The imaginary Brownian particle with the reduced mass, whose motion corresponds to the relative motion of the colliding nuclei, wanders being influenced by the conservative, dissipative, and random (fluctuating) forces. In the present work, only the collisions at the energies well exceeding the Coulomb barrier are considered. That is why the tunneling and channels coupling can be safely neglected.

The motion of the Brownian particle is described by a dimensionless coordinate $q$ which is proportional to the distance between the centers of the colliding nuclei $R$. In [49], it was demonstrated that the orbital degree of freedom could be ignored because it influences the cross-sections within the framework of the statistical errors (typically 1%). In [51], it was proved that, in the collision process, the memory effects at the distances larger than the contact configuration can be discarded. Since in our modeling this configuration is never reached, we use the stochastic Langevin-type equations with the white noise and instant dissipation:

$$dp = \left\{-\frac{dU_{tot}}{dq} + \frac{\hbar^2 L^2}{m_q q^3} - \frac{p}{m_q}K_R\left[\frac{dU_n}{dq}\right]^2\right\}dt + \left|\frac{dU_n}{dq}\right|dW\sqrt{2\theta K_R}, \quad (18)$$

$$dq = \frac{p\,dt}{m_q}. \quad (19)$$

Here $U_{tot} = U_n + U_C$ stands for the total nucleus-nucleus interaction energy; $p$ denotes the linear momentum corresponding to the relative motion of the colliding nuclei; $\hbar L$ is the projection of the orbital angular momentum onto the axis perpendicular to the reaction plane; $K_R$ denotes the dissipation strength coefficient; $\theta$ stands for the thermal energy (temperature). The quantity $m_q$ is the inertia parameter

$$m_q = m_R R_{PT}^2 \quad (20)$$

where $R_{PT} = 1.2\left(A_1^{1/3} + A_2^{1/3}\right)$ fm. In fact, the model is designed in such a way that its physical results are independent on the real value of $R_{PT}$. The time-dependent temperature $\theta$ in Eq. (18) is related to the dissipated energy via the Fermi-gas relation. All details about the way $\theta$ is evaluated can be found in Refs. [13, 49].

The dissipative force (the last term in the figure brackets in Eq. (18)) is related to the nuclear term of the interaction energy via the surface friction expression [50, 52]. The random force (the last term in Eq. (18)) is proportional to the increment of the Wiener process $dW$ which possesses zero average and variance equal to $dt$. Equations (18), (19) are solved numerically using the Runge-Kutta method (see details in [49, 53].

Within the framework of the TM, the transmission coefficient $T_{L\,TM}$ is defined as the number of the captured trajectories divided by the total number of trajectories for each $L$-value. The capture conditions are described in Sec. II F of Ref. [49]. More details about the TM are presented in Refs. [13, 54].

## V. COMPARISON OF THE CROSS SECTIONS WITH THE EXPERIMENT

In Fig. 4, we collect nearly all the above-barrier experimental fusion cross sections $\sigma_{exp}$ found in the literature. For reaction R1 (panel a) five sets of the data are available [40, 41, 43, 55, 56]. Keeping in mind that the sub- and near-barrier data cannot be used in our study, we exclude the data of Refs [57-59] from our consideration. We also omit the data from Refs. [42, 60] because there are only several convenient points in each of these articles and the experimental errors are rather large. One sees that the data in Fig. 4a form two blocks: those from Refs. [40, 55] and from Refs. [41, 43, 56]. Inside each block, the data agree with each other whereas there seems to be some disagreement between the data of different blocks. Therefore, we select the data from [40] and [41] for detailed quantitative separate comparison with the calculations.

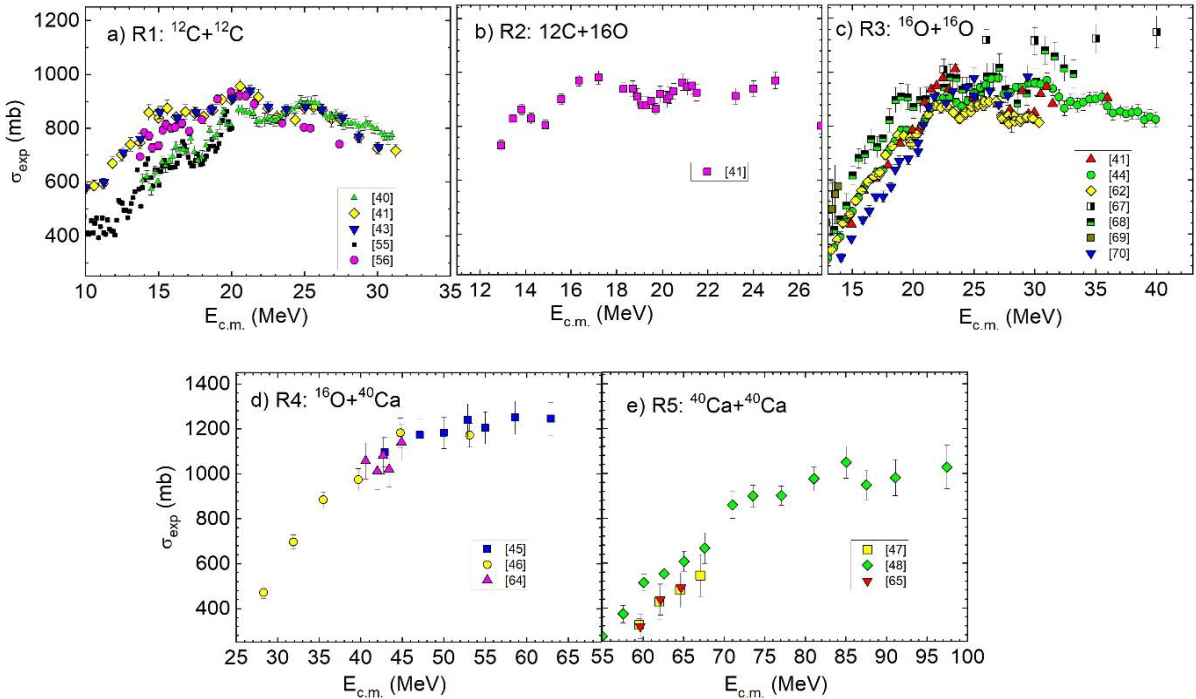

**Fig. 4.** (color online) Experimental fusion cross sections versus the center-of-mass energy. References to the original papers are indicated in the panels. In many cases the data are taken from [61].

Similar problem seems to arise for reaction R3 (see Fig. 4c). However here three data sets [41, 44, 62] are in reasonable agreement with each other. Therefore, we have chosen the single data set [41] for detailed comparison. The non-monotonic oscillating behavior of the cross-sections in reactions R1, R2, R3 was discussed in original experimental papers and in Ref. [63]. Therefore, we do not discuss this behavior here.

Figure 4d shows that the data sets [45, 46, 64] for reaction R4 do not contradict to each other. Below we compare with the calculations the combined data set [46] at lower energies and [45] at higher energies. Finally, for reaction R5 we see in Fig. 4e that data from [48] do not agree with data from [47, 65]. On the other hand, there are only three overlapping points at above barrier energies from [47, 65]. Thus, we select data from [48] and from [47] for separate detailed comparison with calculations. We omit data from [66] since there are only three suitable points with large errors.

For the selected experimental data sets, we perform a quantitative comparison between the theory and experiment using the conventional $\chi^2$ criterium:

$$\chi^2 = \frac{1}{v}\sum_{i=1}^{v}\left(\frac{\sigma_{iTM} - \sigma_{iexp}}{\Delta\sigma_{iexp}}\right)^2. \tag{21}$$

A difficult problem is which collision energies from experimental data set to use for detail quantitative comparison with calculations. In [13, 14] a criterium $\sigma_{exp} > 200$ mb was chosen following Ref. [32]; this approximately corresponds to $E_{c.m.} > 1.1 B_0$. This criterium was formulated in [32] on the basis of the coupled-channels calculations for $^{16}O + ^{154}Sm$ reaction. Accordingly, in [13, 14] the reactions were considered where one of the colliding nuclei was $^{92}Zr$ or heavier. For such relatively heavy nuclei, the energies of the lowest excited states are at most 1 MeV and the barrier energies are about 40 MeV or larger. Considering collision at energies larger than $1.1 B_0$ we safely deal with the situation where at the barrier configuration several excited states are occupied providing dissipation which is essential element of the trajectory model with surface friction.

In the present work, collisions of rather light nuclei are considered with the energy of first excited state being as large as 5 MeV and the barrier energy is as low as 10 MeV. Therefore, even at $E_{c.m.} \approx 2B_0$ we are not surely in the situation when, at the barrier configuration, several excited states are occupied providing dissipation. Therefore, in the present work we apply a phenomenological criterium $E_{c.m.} > E_{cut} = B_0 + E_2$ where $E_2$ is the lowest energy of the second excited state of two colliding nuclei.

Results of the TM-calculations depend upon the friction strength $K_R$ for which in [14] the following empirical formula was obtained

$$K_{Re} = K_1 \exp\left(\frac{B_0 - B_Z}{\Delta B}\right) + K_0. \tag{22}$$

Here

$$B_Z = \frac{Z_1 Z_2}{A_1^{1/3} + A_2^{1/3}} \text{ MeV} \tag{23}$$

is the approximate Coulomb barrier energy.

For reactions $^{12}C+^{12}C$, $^{12}C+^{16}O$, $^{16}O+^{16}O$, Eq. (22) with coefficients of [14] $K_1 = 260$ zs·GeV$^{-1}$, $K_0 = 10$ zs·GeV$^{-1}$, $B_0 = 7$ MeV, $\Delta B = 15$ MeV results in $K_{Re} \approx 190 \div 260$ zs·GeV$^{-1}$. In the present calculations, these values lead to unrealistically small cross sections $\sigma_{TM}$. Yet Eq. (22) as well as its coefficients were based on the cross sections calculated for the reactions with $B_Z > 40$ MeV, whereas for some reactions under consideration in the present work $B_Z < 13$ MeV. Therefore, we attempt using the asymptotic value $K_R = K_0 = 10$ zs·GeV$^{-1}$ as the first approximation and then vary $K_R$ (if possible) trying to reach an agreement with the data.

In Fig. 5a we compare calculated CSs with the experimental ones from Ref. [41] for reaction R1 and from Ref. [44] for reaction R3. The CSs calculated using the barrier penetration model, $\sigma_{BPM}$, significantly overshoot the experimental data at the above barrier collision energies. We consider this as an indication that friction is important ingredient of the process at these energies. It should be noted that the relation $\sigma_{BPM} > \sigma_{exp}$ holds for all considered reactions. The CSs calculated accounting for dissipation, $\sigma_{TM}$, agree with the measured ones much better provided the value of $K_R$ is chosen properly.

Such pictures are typical for other reactions and data sets, that is why we prefer presenting in Fig. 6 the ratios $\sigma_{BPM}/\sigma_{exp}$ and $\sigma_{TM}/\sigma_{exp}$. The values of $\sigma_{TM}$ in these figures correspond to the values of $K_R$ ($K_{Rm}$) providing the minimum value of $\chi^2$ ($\chi_m^2$). The values of $K_{Rm}$ and $\chi_m^2$ as well as the number of data points involved into $\chi^2$-analysis, $v$, are displayed in Table II.

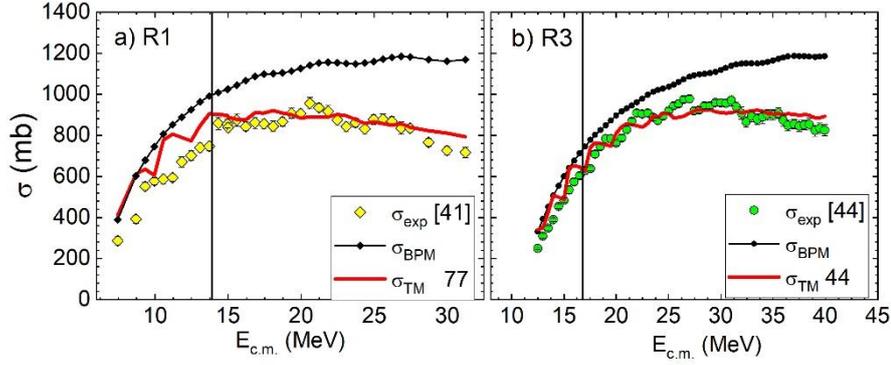

**Fig. 5.** (color online) For two reactions, theoretical fusion cross sections ($\sigma_{BPM}$, lines with symbols; $\sigma_{TM}$, lines without symbols) are compared with the experimental ones (scatter symbols). Digits next to $\sigma_{TM}$ are the values of $K_R$/ (zs · GeV$^{-1}$) (see Table II). Vertical lines correspond to $E_{cut}$.

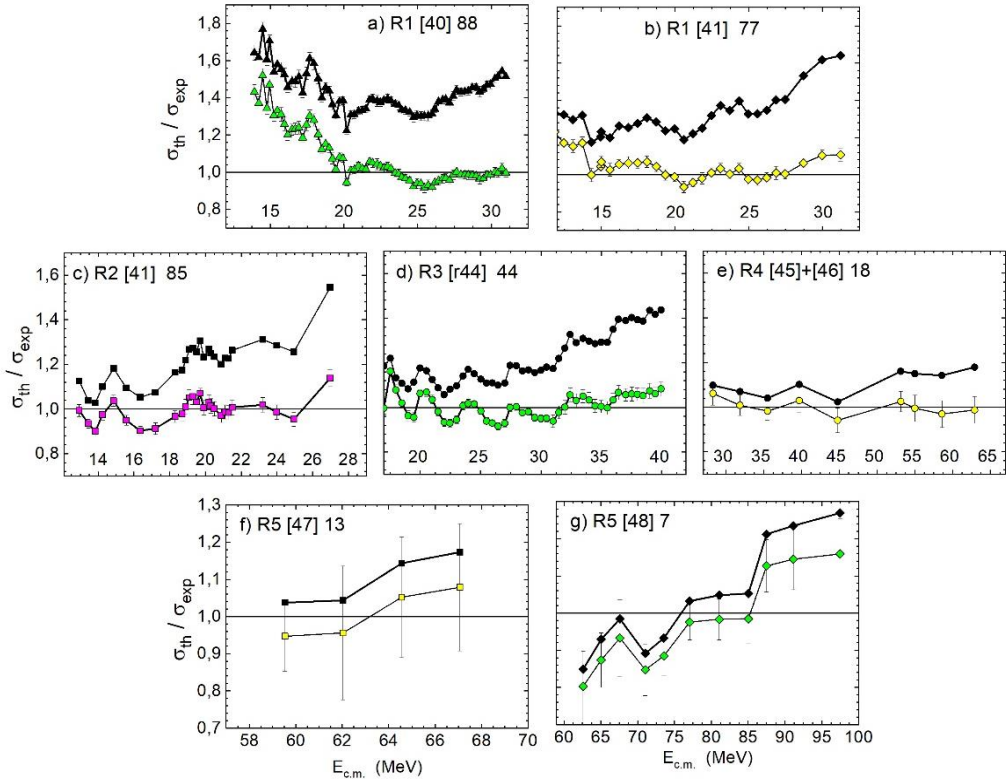

**Fig. 6.** (color online) Ratios $\sigma_{BPM}/\sigma_{exp}$ (upper curves) and $\sigma_{TM}/\sigma_{exp}$ (lower curves) versus collision energy for seven sets of the data selected for detailed quantitative comparison. In each panel, the reaction number, the source of experimental data, and the values of $K_{Rm}/($zs · GeV$^{-1})$ are indicated.

The quantitative results of the comparison between TM and experiment are shown in Table II. One sees that the TM provides rather good description of the data. Note, that at $K_R < 5$ zs · GeV$^{-1}$, the resulting TM cross section becomes insensitive to the value of $K_R$ and $\sigma_{TM}$ becomes very close to $\sigma_{BPM}$.

In Fig. 7, we collect the values of $K_{Rm}$ providing the best fit of the experimental fusion cross sections. Triangles stand for the values of $K_{Rm}$ obtained in Ref. [14]. In that work, the reactions with $B_Z > 40$ MeV were studied, and Eq. (22) was found to fit the values of $K_{Rm}$ with the relative difference equal to 9.0% (this fit is shown in Fig. 7 by thin black line).

Now we have seven more values of $K_{Rm}$ between 7 and 88 zs · GeV$^{-1}$ for $B_Z < 60$ MeV (diamonds in Fig. 7). The new set of $K_{Rm}$ does not really overlaps with the one from Ref. [14] except the heaviest reaction R5 $^{40}$Ca+$^{40}$Ca. The values of $K_{Rm}$ obtained in the present work are well approximated by Eq. (22) with the following set of the coefficients: $K_1 = 80$ zs · GeV$^{-1}$, $K_0 = 10$ zs · GeV$^{-1}$, $B_0 = 7$ MeV, $\Delta B = 8$ MeV (thick blue line in Fig. 7). Although the two curves for $K_{Re}$ differ noticeably, the trend found in [14] holds: the smaller $B_Z$ the larger $K_{Rm}$. The physical meaning of this trend is still unclear for us.

One could think of similarity between the behavior of $K_{Rm}(B_Z)$ and the de Broglie wavelength $\lambda_B(E_{c.m.})$ for the imaginary particle with the reduced mass for which Eqs. (18), (19) are written. We have calculated $\lambda_B(E_{c.m.})$ assuming $Z_1 = Z_2$, $A_1 = 2Z_1 = A_2 = 2Z_2$, and (somewhat arbitrary) $E_{c.m.} = B_Z$. Resulting dependence of $16\lambda_B(B_Z)$ is shown in Fig.7 by pink curve with small boxes. It reproduces the values of $K_{Rm}(B_Z)$ found in the present work amazingly well and probably indicates a way for understanding the physical reason of the $K_{Rm}(B_Z)$-dependence.

Approximation (22) allows us predicting the value $K_{Rm}=24$ zs · GeV$^{-1}$ for reaction $^{12}$C+$^{40}$Ca which is missing in Table II because we did not manage to find corresponding experimental data. Accepting the uncertainty of thus predicted $K_{Rm}$ to be 7 zs · GeV$^{-1}$ we obtain the cross sections shown in Fig.8.

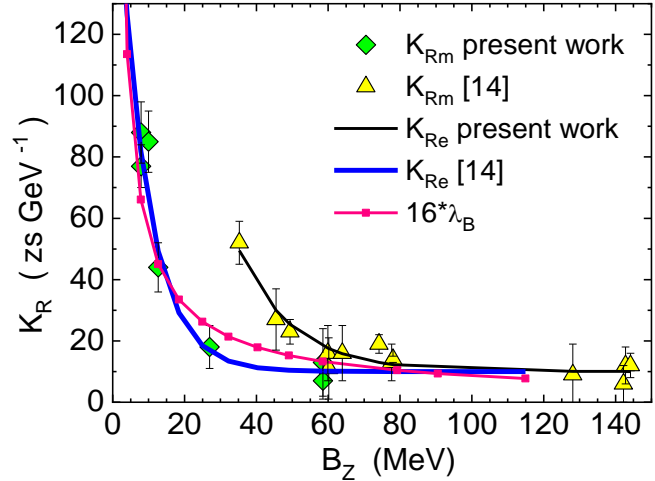

**Fig. 7.** (color online) The values of $K_{Rm}$ obtained in [14] (triangles) and in the present work (diamonds). The lines indicate the approximations using Eq. (22) and the de Broglie wavelength multiplied by 16.

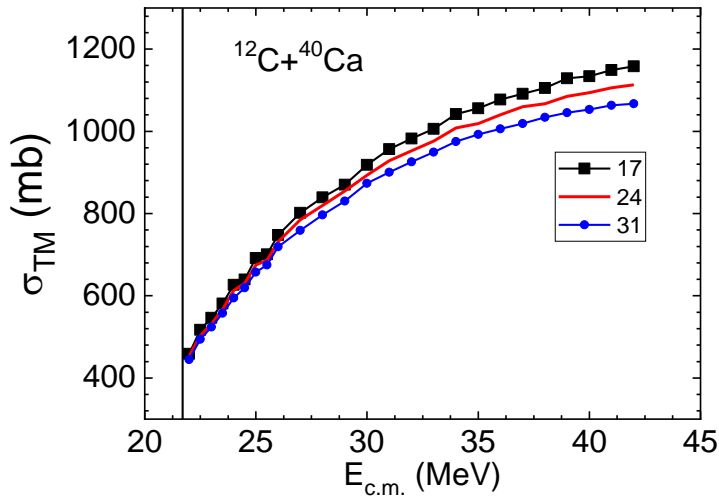

**Fig. 8.** (color online) Predicted fusion cross sections for reaction $^{12}$C+$^{40}$Ca calculated using the three values of $K_R$ indicated in the figure. The vertical line corresponds to $E_{cut}$.

## VI. SUMMARY

One of the methods widely used for obtaining the nucleus-nucleus potential is the semi-microscopical double-folding model. Proton and neutron densities are important ingredient of such a model. The adequate nucleon density (ND) should not only provide good nucleus-nucleus potential but reproduce experimental nuclear charge density as well.

In the present paper, we aimed achieving good description of both the charge nucleon density and the above-barrier fusion cross sections of even-even spherical light nuclei with $Z = N$: $^{12}$C, $^{16}$O, $^{40}$Ca. For this goal, we propose an approximation for the ND for these nuclei. This approximation abbreviated as FE-density (Fermi+exponential) provides good description of the experimental nuclear charge densities.

Then, we use FE nucleon densities for evaluating the nucleus-nucleus potential by means of the double-folding model with the density dependent M3Y-Paris NN-forces. Next these potentials have been applied for calculating the above-barrier fusion cross sections for the reactions $^{12}$C+$^{12}$C, $^{12}$C+$^{16}$O, $^{16}$O+$^{16}$O, $^{16}$O+$^{40}$Ca, and $^{40}$Ca+$^{40}$Ca where the experimental data are available. The cross sections are computed within two approaches: a) the barrier penetration model $\sigma_{BPM}$ and b) the dynamical trajectory model with the surface friction $\sigma_{TM}$. In the latter case, the transmission coefficients are found using the Langevin-type equations.

For all considered reactions, $\sigma_{BPM}$ always exceeds $\sigma_{exp}$. We believe that this is a hint that at the above-barrier energies dissipation plays a significant role in collision process thus requiring a dynamical model [32]. Our trajectory model reproduces the above-barrier experimental cross sections within 10-15%. The only adjustable parameter of this model, the optimal friction strength $K_{Rm}$, was found to be about 90 zs·GeV$^{-1}$ for light reactions $^{12}$C+$^{12}$C, $^{12}$C+$^{16}$O, $^{16}$O+$^{16}$O and about 15 zs·GeV$^{-1}$ for heavier reactions $^{16}$O+$^{40}$Ca and $^{40}$Ca+$^{40}$Ca. This result does not strongly contradict the systematics found earlier in Ref. [14]. Next, we readjust the coefficients of Eq. (22) to build the approximation for the values of friction strength obtained in the present work. Finally, we predict the value of $K_{Rm}$ and calculate the cross sections $\sigma_{TM}$ for reaction $^{12}$C+$^{40}$Ca for which the experimental data seems are missing.

Our FE-algorithm can be applied for above-barrier collision of non-spherical nuclei with $Z = N$ like $^{14}$N, $^{22}$Na, $^{32}$S etc. since it is well known that nuclear deformations do not influence the above-barrier cross sections [32]. We plan to explore this possibility in near future.

**Acknowledgements**

The work was supported by the Foundation for the Advancement of Theoretical Physics and Mathematics "BASIS".